\documentclass[aps,prd,eprint,numerical,showpacs,showkeys,10pt]{revtex4-2}

\usepackage{longtable}
\usepackage{amsmath}
\usepackage{amssymb}
\usepackage{amsfonts}
\usepackage{graphicx}
\usepackage{dcolumn}
\usepackage{bm}
\usepackage{hyperref}
\usepackage{color}
\hypersetup{colorlinks=false}

\usepackage{latexsym,wasysym}

\begin{document}

\title[5DRNBH: Retractions and QBSs]{On the five-dimensional non-extremal Reissner-Nordstr\"{o}m black hole: Retractions and scalar quasibound states}

\date{\today}

\author{M. Abu-Saleem}
\email{mohammedabusaleem2005@yahoo.com; m\_abusaleem@bau.edu.jo}
\affiliation{Mathematics Department, Faculty of Science, Al-Balqa Applied University, 19117 Salt, Jordan}

\author{H. S. Vieira}
\email{Corresponding author: horacio.santana.vieira@hotmail.com; horacio\_vieira@ufla.br; horacio.santana-vieira@tat.uni-tuebingen.de; horacio.santana.vieira@ifsc.usp.br}
\affiliation{Department of Physics, Institute of Natural Sciences, Federal University of Lavras, 37200-000 Lavras, Brazil}
\affiliation{Theoretical Astrophysics, Institute for Astronomy and Astrophysics, University of T\"{u}bingen, 72076 T\"{u}bingen, Germany}
\affiliation{S\~{a}o Carlos Institute of Physics, University of S\~{a}o Paulo, 13560-970 S\~{a}o Carlos, Brazil}

\author{L. H. C. Borges}
\email{luizhenrique.borges@ufla.br}
\affiliation{Department of Physics, Institute of Natural Sciences, Federal University of Lavras, 37200-000 Lavras, Brazil}

\begin{abstract}

In this paper, we examine the role played by topology, and some specific boundary conditions as well, on the physics of a higher-dimensional black hole. We analyze the line element of a five-dimensional non-extremal Reissner-Nordstr\"{o}m black hole to obtain a new family of subspaces that are types of strong retractions and deformations, and then we extend these results to higher dimensions in order to deduce the relationship between various types of transformations. We also study the scalar field perturbations in the background under consideration and obtain an analytical expression for the quasibound state frequencies by using the Vieira-Bezerra-Kokkotas approach, which uses the polynomial conditions of the general Heun functions, and then we discuss the stability of the system and present the radial eigenfunctions. Our main goal is to discuss the physical meaning of these mathematical applications in such higher-dimensional effective metric.

\end{abstract}


%\MSC[2020]: 54C15, 58E10, 81Q05, 83C45, 83C57, 83C75


\preprint{Preprint submitted to Universe/MDPI}

\maketitle

%\begin{quotation}
%...
%\end{quotation}

%
%%%%%%%%%%%%%%%%%%%%%%%%%%%%%%%%%%%%%%%% Introduction
%
\section{Introduction}\label{Introduction}

One of the most exciting areas of physics over the past ten years has been the study of black holes \cite{Beken}, in which it was proposed that rather than being viewed as mathematical constructs, black holes might be thought of as physical systems, and that information theory could be used to comprehend their physics. In order to study quantum gravity, one needs to combine classical mechanics and gravity, and quantum mechanics within the theory of black hole thermodynamics \cite{Bardeen}.

In addition, when combined with a negative cosmological constant, the Reissner--Nordstr\"{o}m black hole (RNBH) has shown to be a valuable tool for research into the thermodynamical features of gravity \cite{Kastor, Kubiz}, as well as to determine the effect of the changed radiation equation of state at the Planck scale on the Reissner--Nordstr\"{o}m--anti-de Sitter black hole (RNAdSBH) \cite{Lobo}. Even though the RNBH in AdS spacetime is a universally efficient solution to many effective field theories of gravity, it still has a lot of baffling problems. It does not saturate the bootstrap bound, even at extremes or at absolute zero, unlike in asymptotically Minkowski spacetime. Without additional charged matter, its classical instability has not been explicitly demonstrated, but it may indicate that the extremal RN solution in AdS spacetime is unstable \cite{Gubser,Konoplya,Gwak}.

It is known that under charged large scalar perturbations, superradiant stability applies to all non-extremal RNBHs in five dimensions. According to this, all RNBHs in higher dimensions, under charged massive scalar perturbations, may be superradiantly stable \cite{Huang1}. On the other hand, the relaxation rate of the four-dimensional RNBH associated with the massless scalar field was analyzed in ref. %MDPI: we revised into ref. please confirm.
 \cite{Hod}.

Asymptotically flat charged black holes can support complex arrangements of both massive and massless scalar fields. These fields exhibit connections to the electromagnetic field tensor of the corresponding central charged black holes (for a review, see the highly fascinating ref. \cite{Herdeiro} and references therein). In the literature \cite{Ishihara,Kodama2008,Ishibashi2003,Konoplya2011,Konoplya2014}, mathematical studies have also been conducted on higher-dimensional RNBHs. A numerical approach was used to demonstrate that asymptotically flat RNBHs, in the range of dimensions $D=(5,6,\ldots,11)$, under massless scalar perturbations, remained relatively stable \cite{Konoplya2009,Huang2}.

The present paper has two broad aims. First, to present some important results on both geometrical and topological aspects of the five-dimensional RNBH spacetime, by constructing different types of strong retractions and strong homotopy retracts on the background under consideration, which is mathematically denoted, for simplicity, by $\mathbf{N}^{5}$ %MDPI: please confirm if the bold in equations need to be retained and if it need to be uniformed in the whole doc. The same to all bold format in all of the equation contents.
 black hole. In addition, we use it to prove the existence of apparent horizons, which are crucial to the studies of wave phenomena on black hole spacetimes. Furthermore, we also show that the end limit of an $\mathbf{N}^{5}$ black hole is a zero-dimensional non-extremal black hole. Second, to discuss the interaction between quantum charged massive scalar particles and the background under consideration, by obtaining the spectrum of quasibound states, which are related to the boundary conditions imposed on the radial wave solution. In addition, we also calculate the corresponding angular and radial wave eigenfunctions. Indeed, these are two very important interesting features from a mathematical physics point of view.

The main interesting field of topology is the theory of strong retracts, and this theory is used to describe the horizon as a subspace in the non-extremal 5DRNBH. In fact, from the theory of geodesics, we infer some types of strong retractions as a cross-sectional apparent horizon of the non-extremal 5DRNBH. Also, quasibound states are wave phenomena that occur near the exterior event horizon of a black hole. However, the relation between the algebraic topology on a non-extremal 5DRNBH and a spectrum of quasibound states comes from the way we can describe the horizons of the non-extremal 5DRNBH in these two different fields.

The quasibound states (or quasistationary levels) are localized in the black hole's potential well, which means that there may exist a flux of particles crossing into the black hole's exterior surface \cite{LettNuovoCim.15.257,RomJPhys.38.729}. Therefore, the spectrum of quasibound states is constituted by complex frequencies, which are denoted by $\omega_{n}=\mbox{Re}[\omega]+i\mbox{Im}[\omega]$, where the imaginary part determines the stability of the system, and the real part is the oscillation frequency. In fact, the wave solution is said to be stable when the imaginary part of the frequency is negative ($\mbox{Im}[\omega] < 0$, decay rate), while the wave solution is said to be unstable when the imaginary part of the frequency is positive ($\mbox{Im}[\omega] > 0$, growth rate). In this work, the quasibound states are obtained by using the so-called Vieira--Bezerra--Kokkotas (VBK) approach \cite{AnnPhys.373.28,PhysRevD.104.024035} (for more details and applications, see also refs. \cite{PhysRevD.105.045015,ModPhysLettA.38.2350160,JHEAp.40.49,PhysRevD.107.104038,PhysLettB.854.138714,EurPhysJC.84.424}).

The paper is organized as follows. In Section \ref{5DRNBH}, we introduce the general metric corresponding to $D$-dimensional RNBH spacetimes, and then we particularize it for the five-dimensional case. In Section \ref{Geometric}, we investigate some geometrical and topological aspects of the $\mathbf{N}^{5}$ black hole and the theory of retracts. In Section \ref{KGE}, we find exact solutions for the charged massive Klein--Gordon equation, and then we obtain an analytical expression for the quasibound states. Finally, in Section \ref{Conclusions}, we summarize the obtained results. {At this point, we utilize} the natural units where $G \equiv c \equiv \hbar \equiv 1$.

%
%%%%%%%%%%%%%%%%%%%%%%%%%%%%%%%%%%%%%%%% Higher-dimensional Reissner-Nordstr\"{o}m black holes
%
\section{Higher-Dimensional Reissner--Nordstr\"{o}m Black Holes}\label{5DRNBH}

The standard four-dimensional black hole solutions to Einstein's equations were generalized to higher dimensions by Myers and Perry in \cite{Myers}. The higher-dimensional action, which is a generalization of the Einstein--Hilbert action, is given by
\begin{equation}
S=\int\biggl(\frac{1}{16 \pi G_{D}}R+\mathcal{L}\biggr)\sqrt{-g}d^{D}x,
\label{eq:action_Myers}
\end{equation}
where $G_{D}$ is the $D$-dimensional gravitational constant, $R$ is the Ricci scalar, $\mathcal{L}$ is the Lagrangian density for any other fields one may wish to consider, and $g \equiv \det(g_{\mu\nu})$. Then, by varying this action with respect to the spacetime metric $g_{\mu\nu}$, we obtain the Einstein equations in higher dimensions
\begin{equation}
R_{\mu\nu}-\frac{1}{2}Rg_{\mu\nu}=8 \pi G_{D} T_{\mu\nu},
\label{eq:Einstein_equations}
\end{equation}
where $R_{\mu\nu}$ is the Ricci tensor. Now, for simplicity and without loss of generality (WLOG), we may adopt the natural units where $G_{D}=1$. Next, by imposing a harmonic gauge condition, a static, asymptotically flat and spherically symmetric vacuum solution of Equation~(\ref{eq:Einstein_equations}), which generalizes the RNBH solution in $D$ dimensions, has the \mbox{following form}
\begin{equation}
ds^{2}=-f(r)dt^{2}+f(r)^{-1} dr^{2}+r^{2} d\Omega_{D-2}^{2},
\label{eq:metric_DDRNBH}
\end{equation}
where the metric function, $f(r)$, and the solid angle element, $d\Omega_{D-2}^{2}$, are given, respectively, by
\begin{equation}
f(r)=1-\frac{2\mathcal{M}}{r^{D-3}}+\frac{\mathcal{Q}^{2}}{r^{2(D-3)}},
\label{eq:metric_function_DDRNBH}
\end{equation}
and %MDPI: please confirm if the indention need to be retained. The same below.
\begin{equation}
d\Omega_{D-2}^{2}=d\theta_{1}^{2}+\sin^{2}\theta_{1} d\theta_{2}^{2}+\cdots+\biggl[\biggl(\prod_{j=1}^{D-3}\sin^{2}\theta_{j}\biggr)d\theta_{D-2}^{2}\biggr].
\label{eq:angle_DDRNBH}
\end{equation}
The parameters $\mathcal{M}$ and $\mathcal{Q}$ are related to the black hole's mass $M$ and charge $Q$, respectively, through the relations
\begin{equation}
\mathcal{M}=\frac{8 \pi}{(D-2)\Omega_{D-2}}M,
\label{eq:mass_DDRNBH}
\end{equation}
and
\begin{equation}
\mathcal{Q}=\frac{8 \pi}{\sqrt{2(D-2)(D-3)}\Omega_{D-2}}Q,
\label{eq:charge_DDRNBH}
\end{equation}
with
\begin{equation}
\Omega_{D-2}=\frac{2\pi^{\frac{D-1}{2}}}{\Gamma(\frac{D-1}{2})},
\label{eq:volume_DDRNBH}
\end{equation}
where $\Omega_{D-2}$ is the volume of a $(D-2)$-dimensional unit sphere, and $\Gamma(x)$ is the gamma function.

Observe that the solution (\ref{eq:metric_DDRNBH}) can be reduced to the standard Reissner--Nordstr\"{o}m metric when $D=4$, as well as to the Schwarzschild one for $\mathcal{Q}=0$. This metric is Ricci flat and is simply called the $D$-dimensional Reissner--Nordstr\"{o}m black hole (DDRNBH) spacetime. {This study concentrates on the }five-dimensional Reissner--Nordstr\"{o}m black hole (5DRNBH). Thus, the explicit line element given by Equation~(\ref{eq:metric_DDRNBH}) for $D=5$ reads
\begin{equation}
ds^{2}=-f(r) dt^{2}+f(r)^{-1} dr^{2}+r^{2}[d\theta^{2}+\sin^{2}\theta(d\phi^{2}+\sin^{2}\phi\ d\chi^{2})],
\label{eq:metric_5DRNBH}
\end{equation}
with %MDPI: we revised equations 10-12 into center alignment, please confirm.
\begin{equation}
f(r)=1-\frac{2\mathcal{M}}{r^{2}}+\frac{\mathcal{Q}^{2}}{r^{4}},
\label{eq:metric_function_5DRNBH}
\end{equation}
\begin{equation}
 \mathcal{M}=\frac{4M}{3\pi}, 
 \label{eq:mass_5DRNBH}
\end{equation}
\begin{equation}
 \mathcal{Q}=\frac{2Q}{\sqrt{3}\pi}, 
 \label{eq:charge_5DRNBH}
\end{equation}
%\begin{eqnarray}
%&& f(r)=1-\frac{2\mathcal{M}}{r^{2}}+\frac{\mathcal{Q}^{2}}{r^{4}}, \label{eq:metric_function_5DRNBH} \\
%&& \mathcal{M}=\frac{4M}{3\pi}, \label{eq:mass_5DRNBH} \\
%&& \mathcal{Q}=\frac{2Q}{\sqrt{3}\pi}, \label{eq:charge_5DRNBH}
%\end{eqnarray}
where $\theta$ and $\phi$ run over the range 0 to $\pi$, and $\chi$ from 0 to $2\pi$. The causal structure of the 5DRNBH spacetime can be identified from the surface equation
\begin{equation}
f(r)=\frac{1}{r^{4}}(r^{4}-2\mathcal{M}r^{2}+\mathcal{Q}^{2})=0=\frac{1}{r^{4}}(r-r_{1})(r-r_{2})(r-r_{3})(r-r_{4}),
\label{eq:surface_equation_5DRNBH}
\end{equation}
whose solutions are the exterior (outer) and interior (inner) ``event'' horizons \cite{Huang3}, given by $r_{1}=\sqrt{\mathcal{M}+\sqrt{\mathcal{M}^{2}-\mathcal{Q}^{2}}}$ and $r_{2}=\sqrt{\mathcal{M}-\sqrt{\mathcal{M}^{2}-\mathcal{Q}^{2}}}$, respectively, and two negative non-physical solutions, given by $r_{3}=-\sqrt{\mathcal{M}-\sqrt{\mathcal{M}^{2}-\mathcal{Q}^{2}}}$ and $r_{4}=-\sqrt{\mathcal{M}+\sqrt{\mathcal{M}^{2}-\mathcal{Q}^{2}}}$.

Then, since the wave phenomena occurring outside the event horizons of a black hole could give us some insight on the physics of these interesting objects, it is meaningful to investigate the existence of such singularities in the background under consideration. Afterward, we discuss a very special type of wave phenomena that occurs near a black hole, namely, the quasibound states of charged massive scalar particles.

\begin{figure}%[H]
%\centering
\includegraphics[width=0.99\columnwidth]{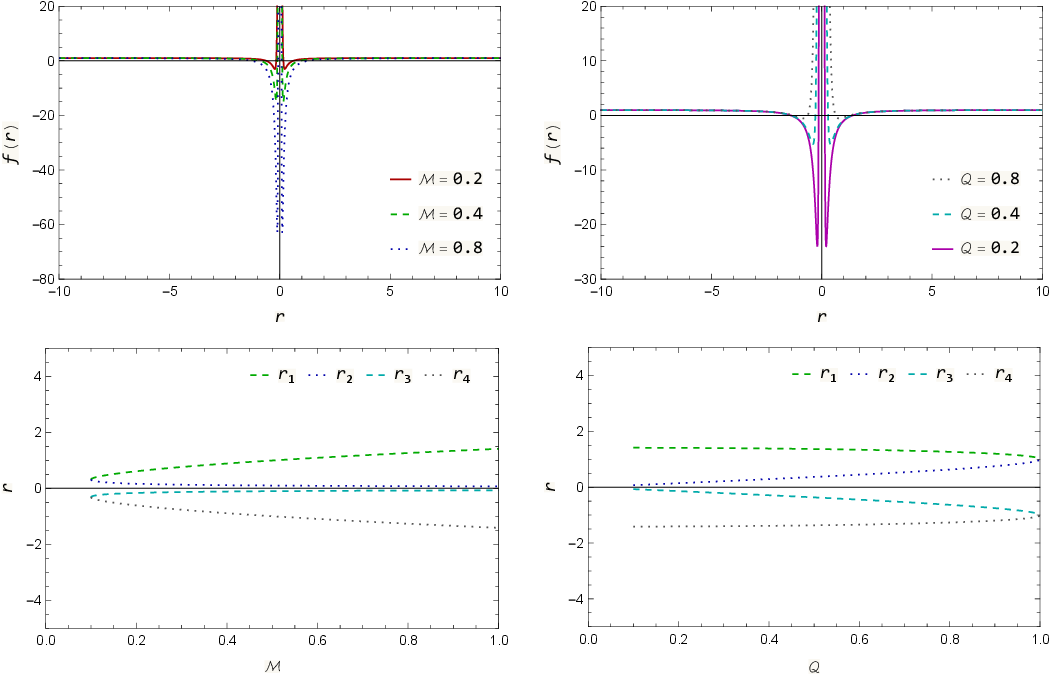}
\caption{Top panel: %MDPI: 1. Please change the hyphen (-) into minus sign ($-$, "U+2212"). e.g., "-1" should be "$-$1". 2. please revise zero into 0. 3. please check if solid line need to add explanation.
 The metric function $f(r)$ with $\mathcal{Q}=0.1$ and varying parameter $\mathcal{M}$ (\textbf{left}), and with $\mathcal{M}=1$ and varying parameter $\mathcal{Q}$ (\textbf{right}). Bottom panel: The horizons with $\mathcal{Q}=0.1$ as functions of the BH's mass $\mathcal{M}$ (\textbf{left}), and with $\mathcal{M}=1$ as functions of the BH's charge $\mathcal{Q}$ (\textbf{right}).}
\label{fig:Fig1_5DRNBH}
\end{figure}

In this work, we concentrate on the non-extreme case, that is, when $\mathcal{M} \neq \mathcal{Q}$. In fact, the 5DRNBH spacetime has apparent horizons (not event horizons, essentially); {these horizons represent the outermost surface where outgoing photons become marginally trapped} ({for a comprehensive investigation} about the radiating black hole horizons, see refs. \cite{Vieira,Saleem1} and references therein). Nevertheless, for simplicity, convenience, and WLOG, we refer to these solutions as the exterior, $r_{1}$, and interior, $r_{2}$, ``event'' horizons. The behavior of the function $f(r)$, as well as that of the horizons, is shown in Figure~\ref{fig:Fig1_5DRNBH}; {Evidently,} the surface equation $f(r)=0$ has four (real) solutions.

%
%%%%%%%%%%%%%%%%%%%%%%%%%%%%%%%%%%%%%%%% Geometric topology and theory of retracts
%
\section{Geometric Topology and Theory of Retracts}\label{Geometric}

{Geometric topology is a field that focuses on exploring the properties of manifolds and maps. Specifically,} we are dealing with embeddings of one manifold into another. Retraction is the continuous mapping from a topological space into a subspace by preserving the position of all points in that subspace, and hence the subspace is called a retract of the original space \cite{Fox}.

In summary, geometric topology concentrates on matters arising in special spaces such as manifolds and the theory of deformations and retractions. If $W$ and $V$ are topological spaces with $W \sqsubseteq V$, then a continuous function $r: V \rightarrow W$ is a retraction if $r(y)=y$ for all $y \in W$. It means that the function $r$ collapses the space $V$ onto its subset $W$. Whenever $W \sqsubseteq V$, the inclusion function $i: W \rightarrow V$ can be defined by $i(y)=y$ within the bigger space $V$. In fact, $r$ is a retraction of $R^{2}-\{0\}$ onto $S^{1}$, since any element $y \in S^{1}$ is sent to itself by $r$ \cite{Fox,Massey,Richter}. As a result, the theory of retracts may be viewed as the concept of determining a class of functions related to topological properties.

In addition to being significant, these maps are more special than arbitrary continuous maps, and these maps are also significantly more generalized than homeomorphisms. Many additional topological aspects are also included in this theory, many of which have a definite geometrical aspect.

Next, we establish two intriguing topological requirements: strong retraction and strong homotopy retraction. After that, {we utilize this theory to analyze the metric of} a $\mathbf{N}^{5}$ black hole (for details, please see ref. \cite{Saleem1}).
%
%%%%%%%%%%%%%%%%%%%%%%%%%%%%%%%%%%%%%%%% Basic definitions
%
\subsection{Basic Definitions}

At this point, we present some essential concepts concerning to the theory of deformations.

\textbf{Definition 1.}
%\noindent \textbf{Definition 1.} 
Let %MDPI: we revised Definition into correct format, please confirm.
 $W$ be an $n$-dimensional manifold with a subspace topology $W_{0} \sqsubseteq W$. A strong retraction is a continuous map $\xi: W \mapsto W_{0}$ for which %\newline
\begin{itemize}
\item[(a)]$W$ %MDPI: We revised into list format, please confirm.
 is open;
\item[(b)]$\xi(y)=y$ $(\forall y \in W_{0})$;
\item[(c)]$\xi(W)=W_{0}$;
\item[(d)]$\xi(W)$ is a manifold with constant curvature %MDPI: please confirm if this is end of definition.
 \cite{Saleem1}.
\end{itemize}

%(a) $W$ is open,% \newline
%(b) $\xi(y)=y$ $(\forall y \in W_{0})$,% \newline
%(c) $\xi(W)=W_{0}$, \newline
%(d) $\xi(W)$ is a manifold with constant curvature \cite{Saleem1}. %\newline

\textbf{Definition 2.}
 A subset $W_{0}$ of a manifold $W$ is called a strong homotopy retract if there exists a strong retraction $\xi: W \mapsto W_{0}$ and a homotopy $\sigma: W \times [0,1] \mapsto W$ for which %\newline
\begin{itemize}
\item[(a)]$\sigma(y,0)=y$ $(\forall y \in W)$;
\item[(b)]$\sigma(y,1)=\xi(y)$ $(\forall y \in W)$;
\item[(c)]$\sigma(y_{0},t)=y_{0}$ $(\forall y_{0} \in W_{0}$ and $\forall t \in [0,1])$ \cite{Saleem1}.
\end{itemize}
%(a) $\sigma(y,0)=y$ $(\forall y \in W)$, %\newline
%(b) $\sigma(y,1)=\xi(y)$ $(\forall y \in W)$, %\newline
%(c) $\sigma(y_{0},t)=y_{0}$ $(\forall y_{0} \in W_{0}$ and $\forall t \in [0,1])$ \cite{Saleem1}. %\newline

\textbf{Definition 3.}
 Suppose that $W_{1}$ and $W_{2}$ are two Riemannian manifolds of dimensions $m_{1}$ and $m_{2}$, respectively. A map $\vartheta: W_{1} \rightarrow W_{2}$ is called an isometric folding of $W_{1}$ into $W_{2}$, if the induced path $\vartheta \circ \gamma: [0,1] \rightarrow W_{2}$ is a piecewise geodesic and of the same length as $\gamma$, whenever a path $\gamma: [0,1] \rightarrow W_{1}$ is a piecewise geodesic \cite{Robertson}. We shall just use the term folding to indicate an isometric folding. In what follows, we introduce the chain of limit folding as $\underset{n \rightarrow \infty}{lim}\vartheta_{n}(W_{n-1})=\underset{n \rightarrow \infty}{lim}\vartheta_{n}(\vartheta_{n-1}(\ldots(\vartheta_{1}(W_{0}))))$, for which $\{\vartheta_{m}: W_{m-1} \rightarrow W_{m}: m=1,2,\ldots,n\}$ is a chain of foldings. If $W_{n}$ is a Riemannian manifold of dimension $n$, and $\underset{n \rightarrow \infty}{lim}\vartheta_{n}(W_{n})=W_{n-1}$, then this type of folding is called \linebreak  folded conditionally.

%
%%%%%%%%%%%%%%%%%%%%%%%%%%%%%%%%%%%%%%%% Components of $\mathbf{N}^{5}$ black hole
%
\subsection{Components of an $\mathbf{N}^{5}$ Black Hole}
In order to investigate some kinds of retractions on the $\mathbf{N}^{5}$ black hole, {we first determine some background parameters and then use the Lagrangian formalism to obtain a set of geodesic equations.}
{Thus, the standard metric for a five-dimensional black hole solution can be written as
\begin{equation}
ds^{2}=-d\mathfrak{p}_{1}^{2}+d\mathfrak{p}_{2}^{2}+d\mathfrak{p}_{3}^{2}+d\mathfrak{p}_{4}^{2}+d\mathfrak{p}_{5}^{2},
\label{eq:general_metric}
\end{equation}
where the components $\mathfrak{p}_{i}$ ($i=1,\ldots,5$) can be constructed by comparing Equations~(\ref{eq:metric_5DRNBH}) and (\ref{eq:general_metric}). Then, by solving (integrating) a system of differential equations, we can find that these components are given by}\vspace{-11pt} 
%\begin{adjustwidth}{-\extralength}{0cm}
%\centering %% If there is a figure in wide page, please release command \centering
\begin{eqnarray}
\mathfrak{p}_{1} & = & \mp \sqrt{\mathrm{1}-\frac{2\mathcal{M}}{r^{2}}+\frac{\mathcal{Q}^{2}}{r^{4}}}t+\mathfrak{a}_{1}, \nonumber\\
\mathfrak{p}_{2} & = & \mp \frac{i\beta_{1}\sqrt{1+\frac{r^{2}}{\beta_{2}}}\sqrt{1-\frac{r^{2}}{\beta_{1}}}\mbox{EllipticE}\biggl[i\sinh^{-1}(\frac{r}{\sqrt{\beta_{2}}}),\frac{\beta_{2}}{\beta_{1}}\biggr]-\mbox{EllipticF}\biggl[i\sinh^{-1}(\frac{r}{\sqrt{\beta_{2}}}),\frac{\beta_{2}}{\beta_{1}}\biggr]}{\frac{r^{2}}{\sqrt{\beta_{2}}}\sqrt{1-\frac{2\mathcal{M}}{r^{2}}+\frac{\mathcal{Q}^{2}}{r^{4}}}}+\mathfrak{a}_{2}, \nonumber\\
\mathfrak{p}_{3} & = & \mp r\theta+\mathfrak{a}_{3}, \nonumber\\
\mathfrak{p}_{4} & = & \mp r(\sin\theta)\phi+\mathfrak{a}_{4}, \nonumber\\
\mathfrak{p}_{5} & = & \mp r(\sin\theta)(\sin\phi)\chi+\mathfrak{a}_{5},
\label{eq11} 
\end{eqnarray}
%\end{adjustwidth}
with
\begin{equation}
\beta_{1}=\mathcal{M}+\sqrt{\mathcal{M}^{2}-\mathcal{Q}^{2}},
\label{beta1}
\end{equation}
\begin{equation}
\beta_{2}=\mathcal{M}-\sqrt{\mathcal{M}^{2}-\mathcal{Q}^{2}},
\label{beta2}
\end{equation}
where $\mathfrak{a}_{i}$ ($i=1,\ldots,5$) are constants (of integration) to be determined. Here, $\mbox{EllipticF}[\mu,\nu]$ and $\mbox{EllipticE}[\mu,\nu]$ are the elliptic integral of the first and second kind, respectively. For simplicity, we can define a five-vector as $\mathfrak{p}=(\mathfrak{p}_{1},\mathfrak{p}_{2},\mathfrak{p}_{3},\mathfrak{p}_{4},\mathfrak{p}_{5})$.
Now, we want to obtain the components of the geodesic equation from the Lagrangian $\mathcal{L}$ and Euler--Lagrange equation, which are given, respectively, by
\begin{equation}
\mathcal{L}=\frac{1}{2}g_{\mu\nu}\dot{\upsilon}^{\mu}\dot{\upsilon}^{\nu},
\label{eq12}
\end{equation}
and
\begin{equation}
\frac{d}{ds}\biggl(\frac{\partial\mathcal{L}}{\partial\dot{\upsilon}^{\mu}}\biggr)-\frac{\partial\mathcal{L}}{\partial\upsilon^{\mu}}=0,
\label{eq13}
\end{equation}
where $\mu,\nu=(1,2,\ldots,5)$, and the dot ``$\ \dot{}\ $'' indicates the derivative with respect to the proper length. Next, by using the relation
\begin{equation}
2\mathcal{L}=ds^{2},
\label{eq14}
\end{equation}
we can write
\begin{equation}
\varpi=-f(r)\dot{t}^{2}+f(r)^{-1}\dot{r}^{2}+r^{2}[\dot{\theta}^{2}+\sin^{2}\theta(\dot{\phi}^{2}+\sin^{2}\phi\  \dot{\chi}^{2}),
\label{eq16}
\end{equation}
where we have used the identity $2\mathcal{L}=\varpi$. %MDPI: please check if \varpi is correct.
 In fact, for null geodesics, we have $\varpi=0$, and for massive particles we get $\varpi=1$. Thus, the components of the geodesic equation are reduced to
\begin{equation}
f(r)\dot{t}=E_{1},
\label{eq17}
\end{equation}
\begin{equation}
-\frac{d}{ds}\biggl(\frac{2\dot{r}}{f(r)}\biggr)+\biggl\{-f'(r)\dot{t}^{2}-\dot{r}^{2}\frac{f'(r)}{[f(r)]^{2}}+2r[\dot{\theta}^{2}+\sin^{2}\theta(\dot{\phi}^{2}+\sin^{2}\phi\ \dot{\chi}^{2})]\biggr\}=0,
\label{eq18}
\end{equation}
\begin{equation}
\frac{d}{ds}(2r^{2}\dot{\theta})-r^{2}[\sin2\theta(\dot{\phi}^{2}+\sin^{2}\phi\ \dot{\chi}^{2})]=0,
\label{eq19}
\end{equation}
\begin{equation}
\frac{d}{ds}(2r^{2}\sin^{2}\theta\ \dot{\phi})-r^{2}(\sin^{2}\theta)(\sin2\phi)\dot{\chi}^{2}=0,
\label{eq20}
\end{equation}
\begin{equation}
r^{2}(\sin^{2}\theta)(\sin^{2}\phi)\dot{\chi}=E_{2},
\label{eq21}
\end{equation}%
where $E_{1}$ and $E_{2}$ are constants (of integration) to be determined, and the quotation marks ``$\ '\ $'' represents the derivative with respect to the radial coordinate $r$.\clearpage

Then, from Equation~(\ref{eq17}), by considering $E_{1}=0$, we obtain a time solution $t=E_{3}$, where $E_{3}$ is also a constant (of integration) to be determined. {Moreover, if we assume that} $f(r)=0$, we have four surfaces, namely, $r_{i}$ ($i=1,\ldots,4$), which are the two event (apparent) horizons and the two non-physical solutions to the surface equation of a 5DRNBH spacetime. Therefore, the surfaces $r_{1}$ and $r_{2}$ describe the exterior and interior event (apparent) horizons of $\mathbf{N}^{5}$. Thus, the photons can escape from the exterior horizon and reach an arbitrary large distance from the black hole; {this confirms that these surfaces are, in fact, apparent horizons, not the event horizons.}

Now, by putting $E_{2}=0$ in Equation~(\ref{eq21}), we obtain at least one of $\theta$, $\phi$, or $\dot{\chi}$ equal to zero. Hence, in the case of $\phi \neq 0$ and $\dot{\chi} \neq 0$, we have $\theta=0$. Thus, we obtain a subspace $\overset{[\theta=0]}{B} \sqsubseteq \mathbf{N}^{5}$, and the components have the following form\vspace{-11pt} 
%\begin{adjustwidth}{-\extralength}{0cm}
%\centering %% If there is a figure in wide page, please release command \centering
\begin{eqnarray}
\overset{[\theta=0]}{\mathfrak{p}_{1}} & = & \mp \sqrt{\mathrm{1}-\frac{2\mathcal{M}}{r^{2}}+\frac{\mathcal{Q}^{2}}{r^{4}}}t+\mathfrak{a}_{1}, \nonumber\\
\overset{[\theta=0]}{\mathfrak{p}_{2}} & = & \mp \frac{i\beta_{1}\sqrt{1+\frac{r^{2}}{\beta_{2}}}\sqrt{1-\frac{r^{2}}{\beta_{1}}}\mbox{EllipticE}\biggl[i\sinh^{-1}(\frac{r}{\sqrt{\beta_{2}}}),\frac{\beta_{2}}{\beta_{1}}\biggr]-\mbox{EllipticF}\biggl[i\sinh^{-1}(\frac{r}{\sqrt{\beta_{2}}}),\frac{\beta_{2}}{\beta_{1}}\biggr]}{\frac{r^{2}}{\sqrt{\beta_{2}}}\sqrt{1-\frac{2\mathcal{M}}{r^{2}}+\frac{\mathcal{Q}^{2}}{r^{4}}}}+\mathfrak{a}_{2}, \nonumber\\
\overset{[\theta=0]}{\mathfrak{p}_{3}} & = & \mathfrak{a}_{3}, \nonumber\\
\overset{[\theta=0]}{\mathfrak{p}_{4}} & = & \mathfrak{a}_{4}, \nonumber\\
\overset{[\theta=0]}{\mathfrak{p}_{5}} & = & \mathfrak{a}_{5}.
\label{eq22} 
\end{eqnarray}
%\end{adjustwidth}
Therefore, we can write the first strong retraction as $\overset{[\theta=0]}{\xi}: \mathbf{N}^{5} \rightarrow \overset{[\theta=0]}{B}$.

Now, by considering $\theta \neq 0$ and $\dot{\chi} \neq 0$, we obtain $\phi =0$, and hence we have a subspace $\overset{[\phi=0]}{B} \sqsubseteq \mathbf{N}^{5}$, whose components are given by\vspace{-11pt}

%\begin{adjustwidth}{-\extralength}{0cm}
%\centering %% If there is a figure in wide page, please release command \centering
\begin{eqnarray}
\overset{[\phi=0]}{\mathfrak{p}_{1}} & = & \mp \sqrt{\mathrm{1}-\frac{2\mathcal{M}}{r^{2}}+\frac{\mathcal{Q}^{2}}{r^{4}}}t+\mathfrak{a}_{1}, \nonumber\\
\overset{[\phi=0]}{\mathfrak{p}_{2}} & = & \mp \frac{i\beta_{1}\sqrt{1+\frac{r^{2}}{\beta_{2}}}\sqrt{1-\frac{r^{2}}{\beta_{1}}}\mbox{EllipticE}\biggl[i\sinh^{-1}(\frac{r}{\sqrt{\beta_{2}}}),\frac{\beta_{2}}{\beta_{1}}\biggr]-\mbox{EllipticF}\biggl[i\sinh^{-1}(\frac{r}{\sqrt{\beta_{2}}}),\frac{\beta_{2}}{\beta_{1}}\biggr]}{\frac{r^{2}}{\sqrt{\beta_{2}}}\sqrt{1-\frac{2\mathcal{M}}{r^{2}}+\frac{\mathcal{Q}^{2}}{r^{4}}}}+\mathfrak{a}_{2}, \nonumber\\
\overset{[\phi=0]}{\mathfrak{p}_{3}} & = & \mp r\theta+\mathfrak{a}_{3}, \nonumber\\
\overset{[\phi=0]}{\mathfrak{p}_{4}} & = & \mathfrak{a}_{4}, \nonumber\\
\overset{[\phi=0]}{\mathfrak{p}_{5}} & = & \mathfrak{a}_{5}.
\label{eq23} 
\end{eqnarray}
%\end{adjustwidth}
Therefore, the strong retraction can be represented as $\overset{[\phi=0]}{\xi}: \mathbf{N}^{5} \rightarrow \overset{[\phi=0]}{B}$.
\clearpage
Next, under the condition that $\theta \neq 0$ and $\phi \neq 0$, we can conclude that $\dot{\chi}=0$, which implies that $\chi=\mbox{constant}=0$ (a special case), and then, there exists a subspace $\overset{[\chi=0]}{B} \sqsubseteq \mathbf{N}^{5}$ that contains the following components\vspace{-11pt}
%\begin{adjustwidth}{-\extralength}{0cm}
%\centering %% If there is a figure in wide page, please release command \centering
\begin{eqnarray}
\overset{[\chi=0]}{\mathfrak{p}_{1}} & = & \mp \sqrt{\mathrm{1}-\frac{2\mathcal{M}}{r^{2}}+\frac{\mathcal{Q}^{2}}{r^{4}}}t+\mathfrak{a}_{1}, \nonumber\\
\overset{[\chi=0]}{\mathfrak{p}_{2}} & = & \mp \frac{i\beta_{1}\sqrt{1+\frac{r^{2}}{\beta_{2}}}\sqrt{1-\frac{r^{2}}{\beta_{1}}}\mbox{EllipticE}\biggl[i\sinh^{-1}(\frac{r}{\sqrt{\beta_{2}}}),\frac{\beta_{2}}{\beta_{1}}\biggr]-\mbox{EllipticF}\biggl[i\sinh^{-1}(\frac{r}{\sqrt{\beta_{2}}}),\frac{\beta_{2}}{\beta_{1}}\biggr]}{\frac{r^{2}}{\sqrt{\beta_{2}}}\sqrt{1-\frac{2\mathcal{M}}{r^{2}}+\frac{\mathcal{Q}^{2}}{r^{4}}}}+\mathfrak{a}_{2}, \nonumber\\
\overset{[\chi=0]}{\mathfrak{p}_{3}} & = & \mp r\theta+\mathfrak{a}_{3}, \nonumber\\
\overset{[\chi=0]}{\mathfrak{p}_{4}} & = & \mp r(\sin\theta)\phi+\mathfrak{a}_{4}, \nonumber\\
\overset{[\chi=0]}{\mathfrak{p}_{5}} & = & \mathfrak{a}_{5}.
\label{eq24} 
\end{eqnarray}
%\end{adjustwidth}
Therefore, with this approach, we can find the other strong retraction, namely, \linebreak  $\overset{[\chi=0]}{\xi}: \mathbf{N}^{5} \rightarrow \overset{[\chi=0]}{B}$.
%
%%%%%%%%%%%%%%%%%%%%%%%%%%%%%%%%%%%%%%%% Strong homotopy retracts
%
\subsection{Strong Homotopy Retracts}

Here, we use all types of strong retractions stated above to introduce strong homotopy retracts of the $\mathbf{N}^{5}$ black hole.

Thus, by using the strong retractions $\overset{[\theta=0]}{\xi}: \mathbf{N}^{5} \rightarrow \overset{[\theta=0]}{B}$, $\overset{[\phi=0]}{\xi}: \mathbf{N}^{5} \rightarrow \overset{[\phi=0]}{B}$, and $\overset{[\chi=0]}{\xi}: \mathbf{N}^{5} \rightarrow \overset{[\chi=0]}{B}$, we can construct the homotopy retracts on $\mathbf{N}^{5}$ as follows.

First, the strong homotopy retracts of $\mathbf{N}^{5}$ onto a geodesic $\overset{[\theta=0]}{B} \sqsubseteq \mathbf{N}^{5}$ can be characterized by
\begin{equation}
\sigma^{\overset{[\theta=0]}{B}}: \mathbf{N}^{5} \times [0,1] \rightarrow N^{5},
\label{eq:strong_theta}
\end{equation}
where
\begin{equation}
\sigma^{\overset{[\theta=0]}{B}}(\mathfrak{p},s)=(1-s)[\sigma^{\overset{[\theta=0]}{B}}(\mathfrak{p},0)]+s[\sigma ^{\overset{[\theta=0]}{B}}(\mathfrak{p},1)],
\label{eq:sigma_theta}
\end{equation}
which is valid $\forall \mathfrak{p} \in \mathbf{N}^{5}$ and $\forall s \in [0,1]$, with
\begin{eqnarray}
\sigma^{\overset{[\theta=0]}{B}}(\mathfrak{p},0) & = & \mathfrak{p}, \nonumber\\
\sigma^{\overset{[\theta=0]}{B}}(\mathfrak{p},1) & = & \biggl\{\biggl(\overset{[\theta=0]}{\mathfrak{p}_{1}},\overset{[\theta=0]}{\mathfrak{p}_{2}},\overset{[\theta=0]}{\mathfrak{p}_{3}},\overset{[\theta=0]}{\mathfrak{p}_{4}},\overset{[\theta=0]}{\mathfrak{p}_{5}}\biggr)\biggr\}.
\label{eq:sigmas_theta}
\end{eqnarray}

Next, we can write the strong homotopy retracts of $\mathbf{N}^{5}$ onto a geodesic $\overset{[\phi=0]}{B} \sqsubseteq \mathbf{N}^{5}$ as
\begin{equation}
\sigma^{\overset{[\phi=0]}{B}}: \mathbf{N}^{5} \times [0,1] \rightarrow N^{5},
\label{eq:strong_phi}
\end{equation}
where
\begin{equation}
\sigma^{\overset{[\phi=0]}{B}}(\mathfrak{p},s)=(1-s)[\sigma^{\overset{[\phi=0]}{B}}(\mathfrak{p},0)]+s[\sigma ^{\overset{[\phi=0]}{B}}(\mathfrak{p},1)],
\label{eq:sigma_phi}
\end{equation}
which is valid $\forall \mathfrak{p} \in \mathbf{N}^{5}$ and $\forall s \in [0,1]$, with
\begin{eqnarray}
\sigma^{\overset{[\phi=0]}{B}}(\mathfrak{p},0) & = & \mathfrak{p}, \nonumber\\
\sigma^{\overset{[\phi=0]}{B}}(\mathfrak{p},1) & = & \biggl\{\biggl(\overset{[\phi=0]}{\mathfrak{p}_{1}},\overset{[\phi=0]}{\mathfrak{p}_{2}},\overset{[\phi=0]}{\mathfrak{p}_{3}},\overset{[\phi=0]}{\mathfrak{p}_{4}},\overset{[\phi=0]}{\mathfrak{p}_{5}}\biggr)\biggr\}.
\label{eq:sigmas_phi}
\end{eqnarray}

Finally, we can achieve the strong homotopy retracts of $\mathbf{N}^{5}$ onto a geodesic $\overset{[\chi=0]}{B} \sqsubseteq \mathbf{N}^{5}$ as
\begin{equation}
\sigma^{\overset{[\chi=0]}{B}}: \mathbf{N}^{5} \times [0,1] \rightarrow N^{5},
\label{eq:strong_chi}
\end{equation}
where
\begin{equation}
\sigma^{\overset{[\chi=0]}{B}}(\mathfrak{p},s)=(1-s)[\sigma^{\overset{[\chi=0]}{B}}(\mathfrak{p},0)]+s[\sigma ^{\overset{[\chi=0]}{B}}(\mathfrak{p},1)],
\label{eq:sigma_chi}
\end{equation}
which is valid $\forall \mathfrak{p} \in \mathbf{N}^{5}$ and $\forall s \in [0,1]$, with
\begin{eqnarray}
\sigma^{\overset{[\chi=0]}{B}}(\mathfrak{p},0) & = & \mathfrak{p}, \nonumber\\
\sigma^{\overset{[\chi=0]}{B}}(\mathfrak{p},1) & = & \biggl\{\biggl(\overset{[\chi=0]}{\mathfrak{p}_{1}},\overset{[\chi=0]}{\mathfrak{p}_{2}},\overset{[\chi=0]}{\mathfrak{p}_{3}},\overset{[\chi=0]}{\mathfrak{p}_{4}},\overset{[\chi=0]}{\mathfrak{p}_{5}}\biggr)\biggr\}.
\label{eq:sigmas_chi}
\end{eqnarray}

In all of these results, Equations~(\ref{eq11}), (\ref{eq22}), (\ref{eq23}), and (\ref{eq24}) provide the values of the components $\mathfrak{p}_{i}$ ($i=1,\ldots,5$).

Therefore, based on these findings, we can infer that all the strong retractions, as a subspace of $\mathbf{N}^{5}$, induce strong homotopy retractions on $\mathbf{N}^{5}$. Furthermore, we deduce that every geodesic provides a strong retraction on a cross-sectional apparent horizon of the 5DRNBH. In addition, the subspaces are strongly homotopically retracted, and they are also contained within a subspace of $\mathbf{N}^{5}$. Thus, we have proved and substantiated (from a rigorous mathematical standpoint) the existence of apparent horizons in the 5DRNBH spacetime. Then, these findings allow us to investigate an interesting physical phenomenon which occurs in the gravity surface of an exterior (apparent) horizon, namely, the quasibound states at the black hole's potential well. Since we have proved the existence of such an exterior horizon, we are able to investigate the boundary condition related to this surface, and at the spatial infinity as well, in order to discuss one of the physical meanings of this mathematical tool. In what follows, first, we conclude the topological studies by investigating the limit case of these transformations, and then we address the motion of scalar particles in the exterior region of the 5DRNBH spacetime by imposing some specific conditions on the radial solution.

%
%%%%%%%%%%%%%%%%%%%%%%%%%%%%%%%%%%%%%%%% Limit transformation on a non-extremal $D$-dimensional RNBH
%
\subsection{Limit Transformation on a Non-Extremal $D$-Dimensional RNBH}

Now, we look at the limit of strong retraction and the limit of folding on a $D$-dimensional non-extremal RNBH from the standpoint of commutative diagrams. 

\textbf{Theorem 1.}
Suppose that $G^{D}$ is a $D$-dimensional non-extremal RNBH. Let $H^{D}$ be a subspace of $G^{D}$. This yields chains of folded conditionally $\vartheta_{i}$ ($i=1,\ldots,n$) and strong retractions $\xi_{j}$ ($j=1,\ldots,n$), with $i,j \in \mathbb{N}$, such that $\vartheta_{1}(G^{D})=G_{1}^{D}$, $\vartheta_{2}(G_{1}^{D})=G_{2}^{D}$, $\ldots$, $\vartheta_{n}(G_{n-1}^{D})=G_{n}^{D}$, in which the end limit is a zero-dimensional non-extremal RNBH ($G^{0}$) with a zero-dimensional subspace ($H^{0}$).

\textbf{Proof.}
In this approach, we introduce the following chains of strong retractions and conditionally folded foldings as commutative diagrams:%Please check intended meaning has been retained
$$
\begin{array}{cccccccc}
G^{D} & \overset{\vartheta_{1}}{\longrightarrow} & G_{1}^{D} & \overset{\vartheta_{2}}{\longrightarrow} & G_{2}^{D} & \cdots & \overset{\underset{n \rightarrow \infty}{\lim} \vartheta_{n}}{\longrightarrow} & G^{D-1} \\ 
\downarrow \xi_{1} & & \downarrow \xi_{2} & & \downarrow \xi_{3} & \cdots &  & \downarrow \underset{n \rightarrow \infty}{\lim} \xi_{n} \\
H^{D} & \overset{\vartheta_{1}}{\longrightarrow} & H_{1}^{D} & \overset{\vartheta_{2}}{\longrightarrow} & H_{2}^{D} & \cdots & \overset{\underset{n \rightarrow \infty}{\lim} \vartheta_{n}}{\longrightarrow} & H^{D-1} \\
\vdots & & \vdots & & \vdots & \cdots & & \vdots \\
G^{D-1} & \overset{\vartheta_{1}}{\longrightarrow} & G_{1}^{D-1} & \overset{\vartheta_{2}}{\longrightarrow} & G_{2}^{D-1} & \cdots & \overset{\underset{n \rightarrow \infty}{\lim} \vartheta_{n}}{\longrightarrow} & G^{D-2} \\ 
\downarrow \xi_{1} & & \downarrow \xi_{2} & & \downarrow \xi_{3} & \cdots & & \downarrow \underset{n \rightarrow \infty}{\lim} \xi_{n} \\
H^{D-1} & \overset{\vartheta_{1}}{\longrightarrow} & \ H_{1}^{D-1} & \overset{\vartheta_{2}}{\longrightarrow} & H_{2}^{D-1} & \cdots & \overset{\underset{n \rightarrow \infty}{\lim} \vartheta_{n}}{\longrightarrow} & H^{D-2} \\
\vdots & & \vdots & & \vdots & \cdots & & \vdots \\ 
G^{1} & \overset{\vartheta_{1}}{\longrightarrow} & G_{1}^{1} & \overset{\vartheta_{2}}{\longrightarrow} & G_{2}^{1} & \cdots & \overset{\underset{n \rightarrow \infty}{\lim} \vartheta_{n}}{\longrightarrow} & G^{0} \\
\downarrow \xi_{1} & & \downarrow \xi_{2} & & \downarrow \xi_{3} & \cdots & & \downarrow \underset{n \rightarrow \infty}{\lim} \xi_{n} \\
H^{1} & \overset{\vartheta_{1}}{\longrightarrow} & H_{1}^{1} & \overset{\vartheta_{2}}{\longrightarrow} & H_{2}^{1} & \cdots & \overset{\underset{n \rightarrow \infty}{\lim} \vartheta_{n}}{\longrightarrow} & H^{0}.
\end{array}%
$$
From the existence of the lower limit, we can consequently obtain the zero-dimensional non-extremal RNBH. 
%$\qquad\qquad\Box$

%
%%%%%%%%%%%%%%%%%%%%%%%%%%%%%%%%%%%%%%%% Charged massive scalar equation of motion
%
\section{Charged Massive Scalar Equation of Motion}\label{KGE}

{In this section, our focus is on the fundamental attributes of the 5DRNBH spacetime,} especially the ones related to their interaction with quantum scalar fields, in particular the classical scalar wave phenomena such as the quasibound states (QBSs). To this end, we consider the covariant, conformally coupled, charged massive Klein--Gordon equation, which is given by
%\begin{adjustwidth}{-\extralength}{0cm}
%\centering %% If there is a figure in wide page, please release command \centering
\begin{equation}
\biggl[\frac{1}{\sqrt{-g}}\partial_{\mu}(g^{\mu\nu}\sqrt{-g}\partial_{\nu})-ie(\partial_{\mu}A^{\mu})-2ieA^{\mu}\partial_{\mu}-\frac{ie}{\sqrt{-g}}A^{\mu}(\partial_{\mu}\sqrt{-g})-e^{2}A^{\mu}A_{\mu}-\mu^{2}\biggr]\Psi=0,
\label{eq:minimally_charged_massive_KG_equation}
\end{equation}
%\end{adjustwidth}
where $g \equiv \mbox{det}(g_{\mu\nu})$, $\Psi=\Psi(t,r,\theta,\phi,\chi)$ is the five-dimensional scalar wave function, and $\mu$ and $e$ are the mass and the charge of the scalar particle, respectively. Here, the electromagnetic five-vector $A_{\mu}$ of the 5DRNBH is given by \cite{PhysLettB.823.136724}
\begin{equation}
A_{\mu}=\frac{\sqrt{3}\mathcal{Q}}{2r^{2}}(-1,0,0,0,0).
\label{eq:electromagnetic_5DRNBH}
\end{equation}
Then, in order to obtain exact analytical solutions of Equation~(\ref{eq:minimally_charged_massive_KG_equation}), due to the stationarity and axisymmetry of the 5DRNBH spacetime, we can use the following ansatz
\begin{equation}
\Psi(t,r,\theta,\phi,\chi)=\mbox{e}^{-i \omega t}U(r)P(\theta,\phi,\chi),
\label{eq:ansatz_5DRNBH}
\end{equation}
where $U(r)=R(r)/r^{3/2}$ is the radial function, $P(\theta,\phi,\chi)$ is an angular function, and $\omega$ is the frequency (energy, in the natural units). Thus, by using the 5DRNBH spacetime metric, the electromagnetic potential, and the separation of variables, given by Equations~(\ref{eq:metric_5DRNBH}), (\ref{eq:electromagnetic_5DRNBH}), and (\ref{eq:ansatz_5DRNBH}), respectively, the equation of motion (\ref{eq:minimally_charged_massive_KG_equation}) can be separated into an angular and a radial equation as
%\begin{adjustwidth}{-\extralength}{0cm}
%\centering %% If there is a figure in wide page, please release command \centering
\begin{equation}
\biggl\{\frac{1}{\sin^{2}\theta}\frac{\partial}{\partial\theta}\biggl(\sin^{2}\theta\frac{\partial}{\partial \theta}\biggr)+\lambda-\frac{1}{\sin^{2}\theta}\biggl[\frac{1}{\sin\phi}\frac{\partial}{\partial\phi}\biggl(\sin\phi\frac{\partial}{\partial\phi}\biggr)\biggr]+\bar{\lambda}+\frac{1}{\sin^{2}\phi}\frac{\partial^{2}}{\partial\chi^{2}}\biggr]\biggr\}P(\theta,\phi,\chi)=0,
\label{eq:angular_5DRNBH}
\end{equation}
%\end{adjustwidth}
and
%\begin{adjustwidth}{-\extralength}{0cm}
%\centering %% If there is a figure in wide page, please release command \centering
\begin{equation}
R^{\prime\prime}(r)+\frac{f^{\prime}(r)}{f(r)}R^{\prime}(r)+\frac{1}{f^{2}(r)}\biggl\{\biggl(\omega-\frac{\varpi_{0}}{r^{2}}\biggr)^{2}-\frac{f(r)[4\lambda+4r^{2}\mu^{2}+3f(r)+6rf^{\prime}(r)}{4r^{2}}\biggr\}R(r)=0,
\label{eq:radial_5DRNBH}
\end{equation}
%\end{adjustwidth}
where $\varpi_{0}=\sqrt{3}\mathcal{Q}e/2$, and $\lambda$ and $\bar{\lambda}$ are the separation constants. Thus, the general exact analytical solutions of the angular equation (\ref{eq:angular_5DRNBH}) are provided by the generic formulas $P(\theta,\phi,\chi)=P_{\nu,4}^{l}(\cos\theta)Y_{lm}(\phi,\chi)$, where $P_{\nu,4}^{l}(\cos\theta)$ are the corresponding Legendre functions in four dimensions \cite{Hochstadt:1973,Frye:2012,Torres:2013} with arbitrary degree $\nu (\in \mathbb{C})$ and order $l$, in which $\lambda=\nu(\nu+2)$, and $Y_{lm}(\phi,\chi)$ are the spherical harmonics, with $l$ and $m$ being the angular and magnetic quantum numbers, respectively, for which $\bar{\lambda}=l(l+1)$. Next, we analytically solve the radial Equation (\ref{eq:radial_5DRNBH}) by applying the VBK approach, which means to rewrite it as a (general) Heun type differential equation \cite{Ronveaux:1995}.

%
%%%%%%%%%%%%%%%%%%%%%%%%%%%%%%%%%%%%%%%% Radial equation
%
\subsection{Radial Equation}

\begin{figure}%[H]
%\centering
\includegraphics[width=1\columnwidth]{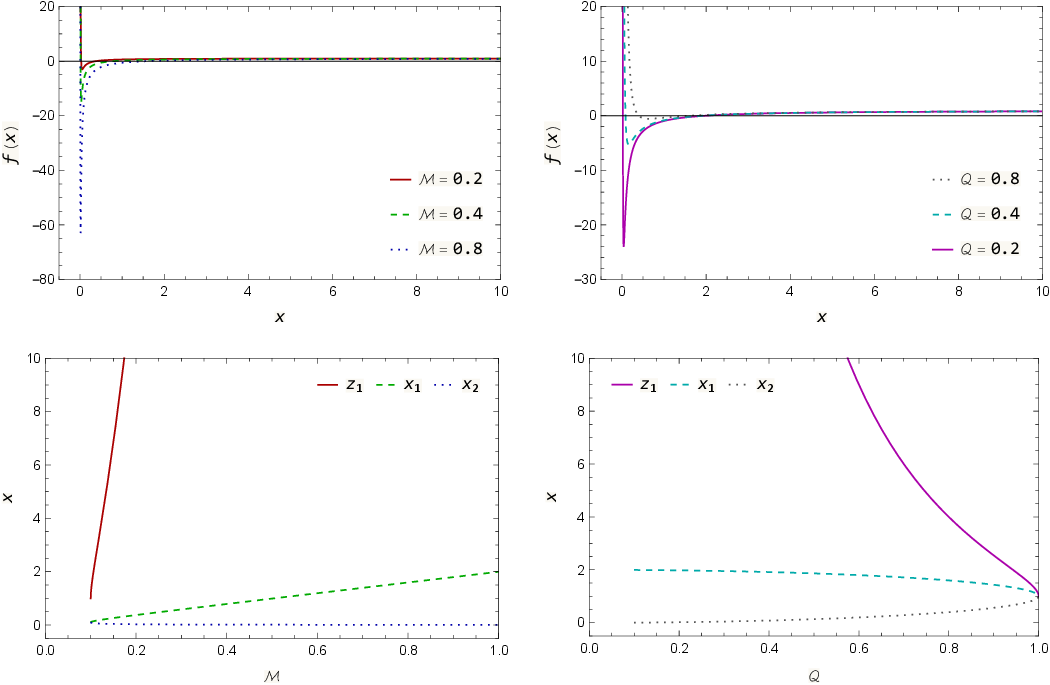}
\caption{Top panel: %MDPI: %MDPI: 1. Please change the hyphen (-) into minus sign ($-$, "U+2212"). e.g., "-1" should be "$-$1". 2. please revise zero into 0. 3. please check if solid line need to add explanation.
 The metric function $f(x)$ with $\mathcal{Q}=0.1$ and varying parameter $\mathcal{M}$ (\textbf{left}), and with $\mathcal{M}=1$ and varying parameter $\mathcal{Q}$ (\textbf{right}). Bottom panel: The horizons with $\mathcal{Q}=0.1$ as functions of the BH's mass $\mathcal{M}$ (\textbf{left}), and with $\mathcal{M}=1$ as functions of the BH's charge $\mathcal{Q}$ (\textbf{right}).}
\label{fig:Fig2_5DRNBH}
\end{figure}

To solve the radial Equation (\ref{eq:radial_5DRNBH}), first, we need to find a more convenient form (parametrization) for the metric function $f(r)$ given by Equation~(\ref{eq:metric_function_5DRNBH}). In fact, the left-hand side of surface Equation (\ref{eq:surface_equation_5DRNBH}) is a biquadratic equation, and we can ``reduce'' the number of singularities by defining a new radial coordinate, $x$, as
\begin{equation}
x=r^{2},
\label{eq:x_coordinate}
\end{equation}
such that
\begin{equation}
f(x)=\frac{1}{x^{2}}(x^{2}-2\mathcal{M}x+\mathcal{Q}^{2}),
\label{eq:metric_function_x_5DRNBH}
\end{equation}
and hence, we get the metric function through parametrization
\begin{equation}
f(x)=0=\frac{1}{x^{2}}(x-x_{1})(x-x_{2}),
\label{eq:surface_equation_x_5DRNBH}
\end{equation}
whose (positive) solutions are now the exterior and interior ``apparent horizons'', given by $x_{1}=\mathcal{M}+\sqrt{\mathcal{M}^{2}-\mathcal{Q}^{2}}$ and $x_{2}=\mathcal{M}-\sqrt{\mathcal{M}^{2}-\mathcal{Q}^{2}}$. Thus, with this parametrized metric function, given by Equation~(\ref{eq:surface_equation_x_5DRNBH}), we demonstrate that Equation~(\ref{eq:radial_5DRNBH}) is totally appropriate to study QBSs with purely ingoing boundary conditions at the exterior event horizon and decaying boundary conditions far from the BH at spatial infinity. The behavior of the metric function $f(x)$, including the horizons, is shown in Figure~\ref{fig:Fig2_5DRNBH}; we can see that the surface equation $f(x)=0$ has two (real, positive) solutions.

Now, by following the steps defined in the VBK approach, we can deduce the exact analytical solutions of Equation~(\ref{eq:radial_5DRNBH}). {First, we need to establish a new radial coordinate}, $z$, as
\begin{equation}
z=\frac{x}{x_{2}},
\label{eq:radial_coordinate_5DRNBH}
\end{equation}
such that
\begin{equation}
z_{1}=\frac{x_{1}}{x_{2}}.
\label{eq:z_1_5DRNBH}
\end{equation}
These new definitions move the three singularities $(0,x_{2},x_{1})$ to the points $(0,1,z_{1})$, as well as keeping a regular singularity at (spatial) infinity. Note that the regular singular point at $z=z_{1}$ is located outside $|z_{1}|>1$. Then, the next step is to perform an \textit{F-homotopic transformation}%MDPI: Please confirm if the italics should be retained.
, such that $R(z) \mapsto y(z)$ by
\begin{equation}
R(z)=z^{\frac{3}{4}}(z-1)^{\frac{\delta-1}{2}}(z-z_{1})^{\frac{\epsilon-1}{2}}y(z),
\label{eq:F-homotopic_5DRNBH}
\end{equation}
where
\begin{eqnarray}
\delta	& = & 1-\frac{i\sqrt{x_{2}}(\omega-\varpi_{2})}{x_{1}-x_{2}},\label{eq:delta_5DRNBH}\\
\epsilon	& = & 1-\frac{i\sqrt{x_{1}}(\omega-\varpi_{1})}{x_{1}-x_{2}},\label{eq:epsilon_5DRNBH}
\end{eqnarray}
with $\varpi_{j}=\varpi_{0}/x_{j}$. Now, by substituting Equations~(\ref{eq:x_coordinate})--(\ref{eq:epsilon_5DRNBH}) into Equation~(\ref{eq:radial_5DRNBH}), we obtain
\begin{equation}
\frac{d^{2} y(z)}{d z^{2}}+\biggl(\frac{\gamma}{z}+\frac{\delta}{z-1}+\frac{\epsilon}{z-z_{1}}\biggr)\frac{d y(z)}{d z}+\frac{\alpha\beta z-q}{z(z-1)(z-z_{1})}y(z)=0,
\label{eq:radial_final_5DRNBH}
\end{equation}
where\vspace{-11pt}
%\begin{adjustwidth}{-\extralength}{0cm}
%\centering %% If there is a figure in wide page, please release command \centering
\begin{eqnarray}
\alpha	& = & \frac{-x_{2} \sqrt{(\mu ^2+4) x_{1}}+x_{1} [\sqrt{(\mu ^2+4) x_{2}}+2 \sqrt{x_{2}}]-i \omega  \sqrt{x_{1} x_{2}}-2 \sqrt{x_{1}} x_{2}+i \varpi_{0}}{2 x_{1} \sqrt{x_{2}}-2 \sqrt{x_{1}} x_{2}},\label{eq:alpha_5DRNBH}\\
\beta	& = & \frac{x_{2} \sqrt{(\mu ^2+4) x_{1}}-x_{1} [\sqrt{(\mu ^2+4) x_{2}}-2 \sqrt{x_{2}}]-i \omega  \sqrt{x_{1} x_{2}}-2 \sqrt{x_{1}} x_{2}+i \varpi_{0}}{2 x_{1} \sqrt{x_{2}}-2 \sqrt{x_{1}} x_{2}},\label{eq:beta_5DRNBH}\\
\gamma	& = & 1,\label{eq:gamma_5DRNBH}\\
q		& = & \frac{1}{4} \biggl[\frac{2 i x_{1} (x_{2} \omega -\varpi_{0})}{x_{2}^{3/2} (x_{2}-x_{1})}-\frac{\varpi_{0}^2}{x_{1} x_{2}^2}-\frac{2 i (x_{1} \omega -\varpi_{0})}{\sqrt{x_{1}} (x_{1}-x_{2})}+\frac{\lambda }{x_{2}}\biggr].\label{eq:q_5DRNBH}
\end{eqnarray}
%\end{adjustwidth}
Equation (\ref{eq:radial_final_5DRNBH}) has the exact (canonical) form of a general Heun equation, where $y(z) \equiv \mbox{HeunG}(z_{1},q;\alpha,\beta,\gamma,\delta;z)$ represents the general Heun function, which is the solution corresponding to the exponent 0 at $z=0$ when considering the value 1 there. %Please check intended meaning has been retained
	Thus, if $\gamma$ is not a negative integer, the existence of $\mbox{HeunG}(z_{1} ,q;\alpha,\beta,\gamma,\delta;z)$ can be deduced from the Fuchs--Frobenius theory, which is analytic in the disk $|z| < 1$ and represented by a Maclaurin series expansion:
\begin{equation}
\mbox{HeunG}(z_{1},q;\alpha,\beta,\gamma,\delta;z)=\sum_{n=0}^{\infty}c_{n}z^{n},
\label{eq:serie_HeunG_todo_x}
\end{equation}
where
\begin{eqnarray}
-qc_{0}+z_{1}  \gamma c_{1} & = & 0,\nonumber\\
P_{n}c_{n-1}-(Q_{n}+q)c_{n}+X_{n}c_{n+1} & = & 0 \quad (n \geq 1),
\label{eq:recursion_General_Heun}
\end{eqnarray}
with
\begin{eqnarray}
P_{n} & = & (n-1+\alpha)(n-1+\beta),\nonumber\\
Q_{n} & = & n[(n-1+\gamma)(1+z_{1} )+z_{1} \delta+\epsilon],\nonumber\\
X_{n} & = & (n+1)(n+\gamma)z_{1} .
\label{eq:P_Q_X_recursion_General_Heun}
\end{eqnarray}
Here, Karl Heun \cite{MathAnn.33.161} adopted the normalization $c_{0}=1$. Also, when $\gamma$ is not a positive integer, the solution of Equation~(\ref{eq:radial_final_5DRNBH}) corresponding to the exponent $1-\gamma$ at $z=0$ is $z^{1-\gamma}\mbox{HeunG}(z_{1},(z_{1} \delta+\epsilon)(1-\gamma)+q;\alpha+1-\gamma,\beta+1-\gamma,2-\gamma,\delta;z)$.

{Hence}, the exact solution for the radial part of the conformally coupled charged massive Klein--Gordon equation in the 5DRNBH spacetime can be represented as
\begin{equation}
R_{j}(z)=z^{\frac{3}{4}}(z-1)^{\frac{\delta-1}{2}}(z-z_{1})^{\frac{\epsilon-1}{2}}[C_{1,j}\ y_{1,j}(z) + C_{2,j}\ y_{2,j}(z)],
\label{eq:analytical_solution_radial_5DRNBH}
\end{equation}
where $C_{1,j}$ and $C_{2,j}$ are constants (to be determined), and $j=\{0,1,z_{1},\infty\}$ labels the solution at all singular points.  {Consequently, we obtain a pair of linearly independent solutions at $z=0$ ($r=0$) through}\vspace{-11pt}
%\begin{adjustwidth}{-\extralength}{0cm}
%\centering %% If there is a figure in wide page, please release command \centering
\begin{eqnarray}
y_{1,0}(z) & = & \mbox{HeunG}(z_{1},q;\alpha,\beta,\gamma,\delta;z),\label{eq:y10_5DRNBH}\\
y_{2,0}(z) & = & z^{1-\gamma}\mbox{HeunG}(z_{1},(z_{1} \delta+\epsilon)(1-\gamma)+q;\alpha+1-\gamma,\beta+1-\gamma,2-\gamma,\delta;z).\label{eq:y20_5DRNBH}
\end{eqnarray}
%\end{adjustwidth}
Similarly, the pair of linearly independent solutions for exponents $0$ and $1-\delta$ at $x=1$ ($r=r_{2}$) is provided by\vspace{-11pt}
%\begin{adjustwidth}{-\extralength}{0cm}
%\centering %% If there is a figure in wide page, please release command \centering
\begin{eqnarray}
y_{1,1}(z) & = & \mbox{HeunG}(1-z_{1} ,\alpha\beta-q;\alpha,\beta,\delta,\gamma;1-z),\label{eq:y11_5DRNBH_1}\\
y_{2,1}(z) & = & (1-z)^{1-\delta}\mbox{HeunG}(1-z_{1} ,((1-z_{1} )\gamma+\epsilon)(1-\delta)+\alpha\beta-q;\alpha+1-\delta,\beta+1-\delta,2-\delta,\gamma;1-z).\label{eq:y21_5DRNBH_1}
\end{eqnarray}
%\end{adjustwidth}
The pair of linearly independent solutions for exponents $0$ and $1-\epsilon$ at $z=z_{1} $ ($r=r_{1}$) is provided by%\vspace{-11pt}

%\begin{adjustwidth}{-\extralength}{0cm}
%\centering %% If there is a figure in wide page, please release command \centering
{\small
\begin{eqnarray}
y_{1,z_{1}}(z) & = & \mbox{HeunG}\biggl(\frac{z_{1} }{z_{1} -1},\frac{\alpha\beta z_{1} -q}{z_{1} -1};\alpha,\beta,\epsilon,\delta;\frac{z_{1} -z}{z_{1} -1}\biggl),\label{eq:y1x1_5DRNBH_1}\\
y_{2,z_{1}}(z) & = & \biggl(\frac{z_{1} -z}{z_{1} -1}\biggl)^{1-\epsilon}\mbox{HeunG}\biggl(\frac{z_{1} }{z_{1} -1},\frac{(z_{1} (\delta+\gamma)-\gamma)(1-\epsilon)}{z_{1} -1}+\frac{\alpha\beta z_{1} -q}{z_{1} -1};\alpha+1-\epsilon,\beta+1-\epsilon,2-\epsilon,\delta;\frac{z_{1} -z}{z_{1} -1}\biggl).\label{eq:y2x1_5DRNBH_1}
\end{eqnarray}}
%\end{adjustwidth}
Finally, the pair of linearly independent solutions for exponents $\alpha$ and $\beta$ at $z=\infty$ ($r=\infty$) is provided by%\vspace{-11pt}

%\begin{adjustwidth}{-\extralength}{0cm}
%\centering %% If there is a figure in wide page, please release command \centering
\begin{eqnarray}
y_{1,\infty}(z) & = & z^{-\alpha}\mbox{HeunG}\biggl(\frac{1}{z_{1} },\alpha(\beta-\epsilon)+\frac{\alpha}{z_{1} }(\beta-\delta)-\frac{q}{z_{1} };\alpha,\alpha-\gamma+1,\alpha-\beta+1,\delta;\frac{1}{z}\biggl),\label{eq:y1i_5DRNBH_1}\\
y_{2,\infty}(z) & = & z^{-\beta}\mbox{HeunG}\biggl(\frac{1}{z_{1} },\beta(\alpha-\epsilon)+\frac{\beta}{z_{1} }(\alpha-\delta)-\frac{q}{z_{1} };\beta,\beta-\gamma+1,\beta-\alpha+1,\delta;\frac{1}{z}\biggl).\label{eq:y2i_5DRNBH_1}
\end{eqnarray}
%\end{adjustwidth}

Next, by assuming a specific asymptotic behavior (boundary condition) on these exact solutions, namely, near the exterior black hole event horizon ($r \rightarrow r_{1}$) and at the spatial infinity ($r \rightarrow \infty$), we can discuss the physical solutions related to the quasibound states.
%
%%%%%%%%%%%%%%%%%%%%%%%%%%%%%%%%%%%%%%%% Quasibound states
%
\subsection{Quasibound States}

Now, we apply the VBK approach \cite{AnnPhys.373.28,PhysRevD.104.024035} to deduce the characteristic resonance equation and then determine the spectrum of QBS frequencies. To do this, we need to impose the following two boundary conditions on the radial solution: it should describe a purely ingoing wave at the exterior black hole event horizon and tend (decaying) to zero far from the black hole at spatial infinity. Since QBSs are mode solutions to the eigenvalue problem, we refer to QBSs as $\omega_{n}$. In addition, it can be shown that the fact that at spatial infinity QBSs decay exponentially enables an analytic evaluation. Here, we only present the most important and final results, and hence we cordially invite the readers to check all the necessary calculations/steps on the VBK approach papers \cite{AnnPhys.373.28,PhysRevD.104.024035}.

Thus, the general Heun function becomes a (class I) polynomial of degree $n$ $(\geq 0)$ iff it satisfies two conditions \cite{Ronveaux:1995}, namely,
\begin{eqnarray}
\alpha+n   & = & 0,\label{eq:1st_condition}\\
c_{n+1}(q) & = & 0.\label{eq:2nd_condition}
\end{eqnarray}
From the first condition, which is called $\alpha$-condition, we find the frequency eigenvalues. Moreover, based on the second condition, we determine the constant $\lambda$ for all value of $n$,  and then we use it to find the eigenvalues $\nu$, as well as the radial and angular wave eigenfunctions. Therefore, by implementing  Equation~(\ref{eq:1st_condition}), we have that the exact analytical spectrum of QBSs is provided by
\begin{equation}
\omega_{n}=\frac{\varpi_{0}+i\{x_{2}[2(n+1)\sqrt{x_{1}}+\sqrt{(\mu^{2}+4)x_{1}}]-x_{1}[2(n+1)\sqrt{x_{2}}+\sqrt{(\mu^{2}+4)x_{2}}]\}}{\sqrt{x_{1}x_{2}}},
\label{eq:omega_5DRNBH}
\end{equation}
where $n$ is now the overtone number, which can be called the principal quantum number. In fact, $\mbox{Re}[\omega]=\varpi_{0}$. On the other hand, by imposing Equation~(\ref{eq:2nd_condition}) for the fundamental mode when $n=0$ leads to
\begin{equation}
c_{n+1}(q)\biggl|_{n=0}=q_{0}=q=0,
\label{eq:q_0_5DRNBH}
\end{equation}
whose solution is
\begin{equation}
\lambda_{0}=\frac{\varpi_{0}^{2}}{x_{1}x_{2}}+\frac{2i[x_{1}x_{2}\omega-\varpi_{0}(x_{1}+\sqrt{x_{1}x_{2}}-x_{2})]}{\sqrt{x_{1}x_{2}}(\sqrt{x_{1}}-\sqrt{x_{2}})},
\label{eq:lambda_0_5DRNBH}
\end{equation}
from which we can determine the angular eigenvalue $\nu_{0}$ for the ground state, as, for example, $\nu_{0}=1.25178-0.03845i$ for $\mathcal{M}=1$, $\mu=0.5$, $\mathcal{Q}=0.5$, and $e=0.1$.

\begin{figure}%[H]
%\centering
\includegraphics[width=1\columnwidth]{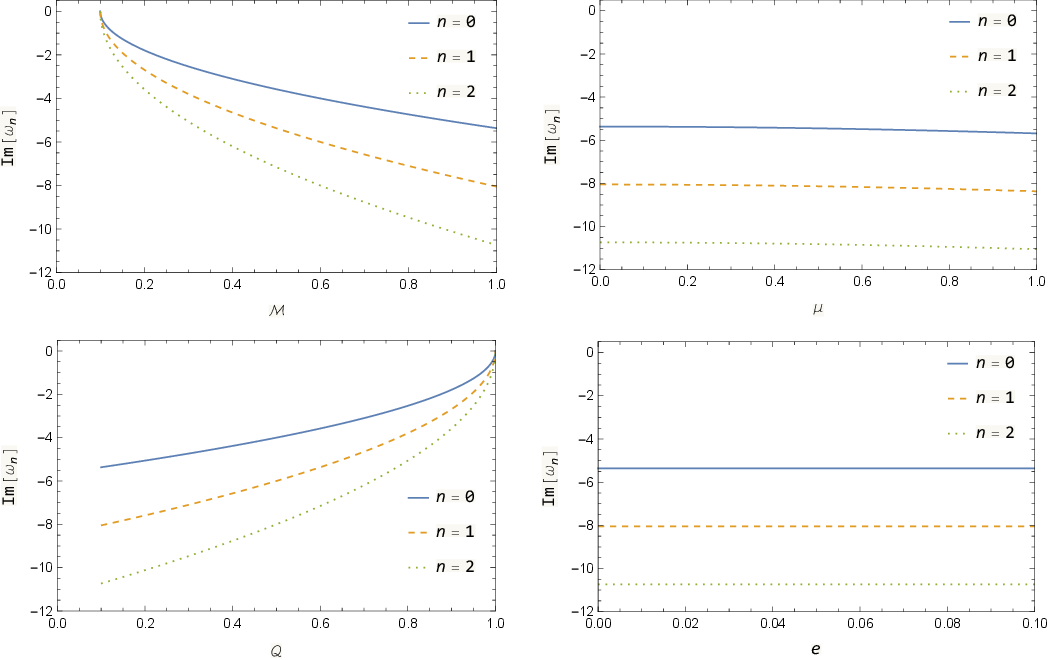}
\caption{Top panel: %MDPI: %MDPI: %MDPI: 1. Please change the hyphen (-) into minus sign ($-$, "U+2212"). e.g., "-1" should be "$-$1". 2. please revise zero into 0.
 The imaginary part of the charged massive scalar QBSs $\omega_{n}$ of a five-dimensional non-extremal Reissner--Nordstr\"{o}m black hole with respect to $n=0,1,2$ and varying the BH's mass $\mathcal{M}$ (\textbf{left}) and the field's mass $\mu$ (\textbf{right}). Bottom panel: The imaginary part of the charged massive scalar QBSs $\omega_{n}$ of a five-dimensional non-extremal Reissner--Nordstr\"{o}m black hole with respect to $n=0,1,2$ and varying the BH's charge $\mathcal{Q}$ (\textbf{left}) and the field's charge $e$ (\textbf{right}).}
\label{fig:Fig3_5DRNBH}
\end{figure}

In Figure~\ref{fig:Fig3_5DRNBH} we show the behavior of the imaginary part of charged massive scalar QBSs in the 5DRNBH spacetime; The real part for the QBS $\omega_{n}$ is always constant and equal to $\varpi_{0}$.

%
%%%%%%%%%%%%%%%%%%%%%%%%%%%%%%%%%%%%%%%%%%%%%%%%%%%%%%%%%%%%%%%%% Radial wave eigenfunctions
%
\subsection{Radial Wave Eigenfunctions}\label{RWE}
Finally, we introduce the radial wave eigenfunctions, which correspond to the charged massive scalar QBS frequencies in the 5DRNBH spacetime. To do this, we also follow the VBK approach.

Thus, the QBS radial wave eigenfunctions for charged massive scalar fields propagating in a 5DRNBH spacetime are given by
\begin{equation}
U_{n}(z) = \frac{1}{z^{\frac{3}{4}}} R(z) = C_{n} (z-1)^{\frac{\delta-1}{2}}(z-z_{1})^{\frac{\epsilon-1}{2}} \mbox{HeunGp}_{n}(z_{1},q_{n};-n,\beta,\gamma,\delta;z),
\label{eq:RWE_4DUABH}
\end{equation}
where $C_{n}$ is a constant to be determined, and $\mbox{HeunGp}_{n}(z_{1},q_{n};-n,\beta,\gamma,\delta;z)$ are the general Heun polynomials.

\begin{figure}%[H]
%\centering
\includegraphics[width=1\columnwidth]{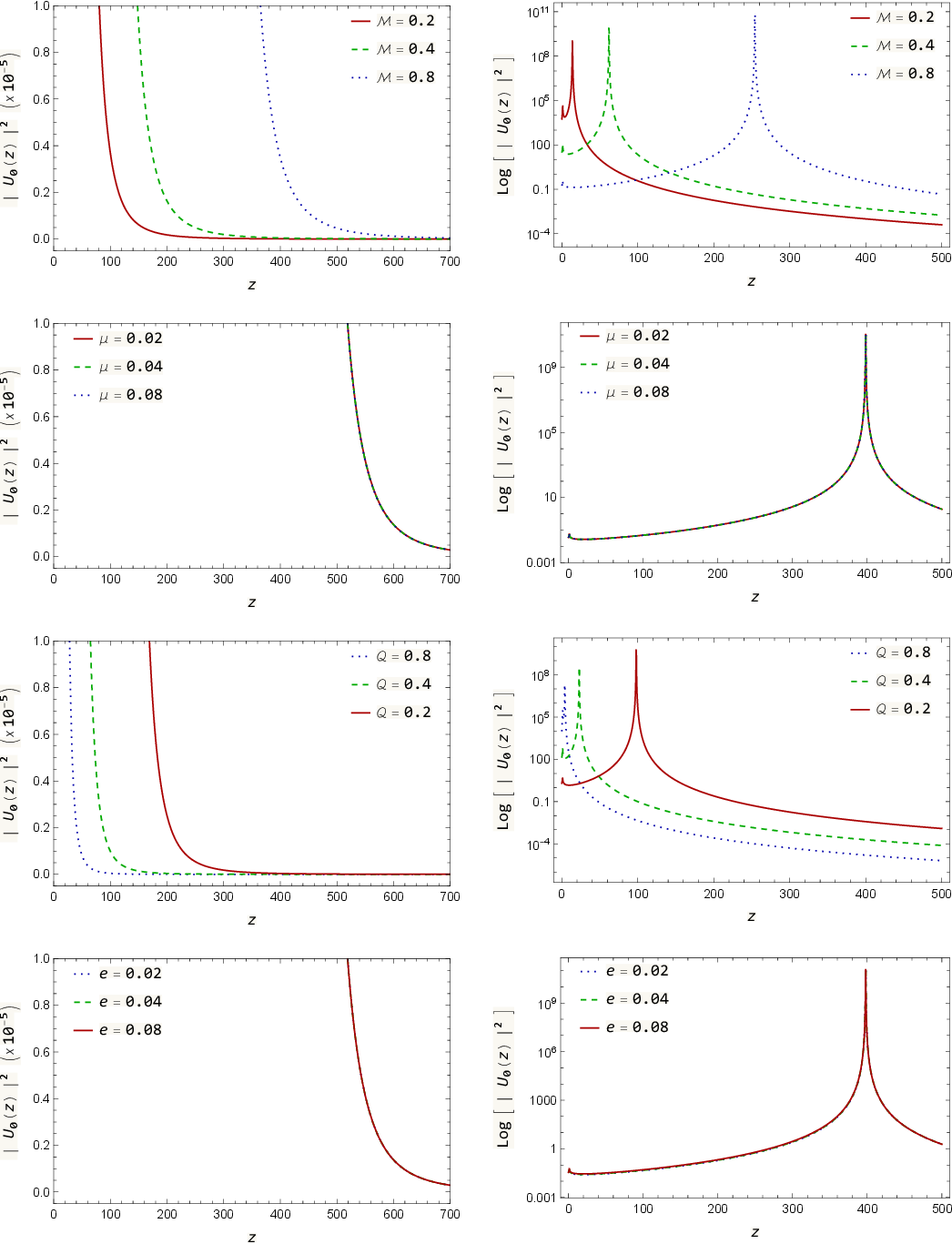}
\caption{The %MDPI: %MDPI: %MDPI: %MDPI: 1. Please change the hyphen (-) into minus sign ($-$, "U+2212"). e.g., "-1" should be "$-$1". 2. please revise zero into 0. 3. some lines are overlapping, please check if this is ok.
 squared radial wave eigenfunctions (\textbf{left}) and their log-scale plots (\textbf{right}) for the fundamental $n=0$ of charged massive scalar QBS frequencies $\omega_{n}$ in a 5DRNBH spacetime as functions of $\mathcal{M}$, $\mu$, $\mathcal{Q}$, and $e$. The units are in multiples of $C_{n}$.}
\label{fig:Fig4_5DRNBH}
\end{figure}

In Figure~\ref{fig:Fig4_5DRNBH}, we present the squared radial wave eigenfunctions for the fundamental mode. We can see that these radial solutions tend to zero at spatial infinity and diverge (reaching a maximum value) at the exterior event horizon.

%
%%%%%%%%%%%%%%%%%%%%%%%%%%%%%%%%%%%%%%%% Final remarks
%
\section{Final Remarks}\label{Conclusions}

{In this work, we investigated novel mathematical implications of concepts in geometric topology, retract theory, and physics, focusing on strong homotopy retracts in the five-dimensional non-extremal Reissner--Nordstr\"{o}m black hole spacetime}. From the metric describing the aforementioned background, we showed that the geodesics were strong retractions. In addition, we also proved that the end limit of folding on a $D$-dimensional non-extremal Reissner--Nordstr\"{o}m black hole was a zero-dimensional non-extremal RNBH.

We used the theory of retracts to investigate the importance of geometric topology in the study of spacetime singularities, especially the apparent horizons of the five-dimensional non-extremal Reissner--Nordstr\"{o}m black hole.
We also looked at how quantum charged massive scalar fields behaved as they moved near the exterior event horizon of the five-dimensional non-extremal Reissner--Nordstr\"{o}m black hole spacetime. We deduced the exact analytical solution for the radial part of the conformally coupled Klein--Gordon equation, provided in terms of the general Heun functions, and subsequently studied a significant physical phenomenon associated with the boundary conditions imposed on it, namely, the quasibound states. The spectrum of quasibound state frequencies for charged massive scalar fields was obtained by using the VBK approach.

Finally, we discussed the stability of the system. The system was stable, since the imaginary part of its resonant frequencies were negative. Here, it is worth emphasizing that these quasibound state frequencies were obtained directly from the general Heun functions by using their polynomial conditions, and, to our knowledge, there have not been any similar results in the literature for the background under consideration. In fact, there exist some results on bound states (not quasibound) of both 4D and 5D RNBHs obtained by Huang, Zhao ,and Zou \cite{PhysLettB.823.136724}, where they imposed a bound state condition given by $\omega^{2} < \mu^{2}$, which is a low-frequency limit, meaning that the particle was not absorbed by the black hole, and eventually, it could be emitted with a superradiant frequency; therefore, the spectrum of eigenfrequencies was just real, that is, $\omega=\mbox{Re}[\omega]$, such that it did not depend on an overtone number $n$. %Please check intended meaning has been retained
Furthermore, in our case, the radial solution diverged near the exterior event horizon, and achieving a maximum value indicated that the scalar particles could cross into the \linebreak  black hole.

We hope that, in the near future, our results can be used to fit some astrophysical and/or experimental data and hence may shed some light on the physics of black holes and higher-dimensional theories of gravity as well.

%
%%%%%%%%%%%%%%%%%%%%%%%%%%%%%%%%%%%%%%%%%%%%%%%%%%%%%%%%%%%%%%%%%%%%%%%%%%%%%%%%%%%%%%%%%%%%%% Author contributions
%
%\section*{Author contributions}
%H. S. Vieira: Conceptualization, investigation, methodology, software, writing.
%
%K. D. Kokkotas: Project administration, supervision, validation.
%
%%%%%%%%%%%%%%%%%%%%%%%%%%%%%%%%%%%%%%%%%%%%%%%%%%%%%%%%%%%%%%%%%%%%%%%%%%%%%%%%%%%%%%%%%%%%%% Declarations
%
\section*{Declarations}
%
%%%%%%%%%%%%%%%%%%%%%%%%%%%%%%%%%%%%%%%%%%%%%%%%%%%%%%%%%%%%%%%%%%%%%%%%%%%%%%%%%%%%%%%%%%%%%% Data Availability Statement
%
\subsection*{Data Availability Statement}

The data that support the findings of this study are available from the corresponding author upon reasonable request.

%
%%%%%%%%%%%%%%%%%%%%%%%%%%%%%%%%%%%%%%%%%%%%%%%%%%%%%%%%%%%%%%%%%%%%%%%%%%%%%%%%%%%%%%%%%%%%%% Acknowledgments
%
\subsection*{Acknowledgments}

H.S.V. is partially supported by the Alexander von Humboldt-Stiftung/Foundation (Grant No. 1209836). This study was financed in part by the Conselho Nacional de Desenvolvimento Cient\'{i}fico e Tecnol\'{o}gico -- Brasil (CNPq) -- Research Project No. 150410/2022-0. Funded by the Federal Ministry of Education and Research (BMBF) and the Baden-W\"{u}rttemberg Ministry of Science as part of the Excellence Strategy of the German Federal and State Governments -- Reference No. 1.-31.3.2/0086017037. It is a great pleasure to thank the Theoretical Astrophysics at T\"{u}bingen (TAT Group) for its wonderful hospitality and technological support (software licenses).

%
%%%%%%%%%%%%%%%%%%%%%%%%%%%%%%%%%%%%%%%%%%%%%%%%%%%%%%%%%%%%%%%%%%%%%%%%%%%%%%%%%%%%%%%%%%%%%% Conflicts of Interest
%
\subsection*{Conflicts of Interest}

The authors declare they have no financial interests. The authors declare they have no conflict of interest. The funders had no role in the design of the study; in the collection, analyses, or interpretation of data; in the writing of the manuscript; or in the decision to publish the results.

%
%%%%%%%%%%%%%%%%%%%%%%%%%%%%%%%%%%%%%%%%%%%%%%%%%%%%%%%%%%%%%%%%%%%%%%%%%%%%%%%%%%%%%%%%%%%%%% Competing interests
%
\subsection*{CRediT authorship contribution statement}

M. Abu-Saleem: Conceptualization, Methodology, Investigation, Writing - original draft. H. S. Vieira: Software, Data curation, Visualization, Supervision, Validation, Writing - review \& editing. L. H. C. Borges: Supervision, Validation, Writing - review \& editing.

%
%%%%%%%%%%%%%%%%%%%%%%%%%%%%%%%%%%%%%%%%%%%%%%%%%%%%%%%%%%%%%%%%%%%%%%%%%%%%%%%%%%%%%%%%%%%%%% thebibliography
%

%
%%%%%%%%%%%%%%%%%%%%%%%%%%%%%%%%%%%%%%%%%%%%%%%%%%%%%%%%%%%%%%%%%%%%%%%%%%%%%%%%%%%%%%%%%%%%%%
%
\end{document}